\documentclass[a4paper,11pt]{article}
\pdfoutput=1 
\usepackage{jcappub}
\usepackage{xcolor}
\usepackage{lineno}
\usepackage{soul}
\usepackage[normalem]{ulem}

\usepackage[T1]{fontenc} 

\newcommand{\CH}[1]{{\color{black}{{#1}}}}

\title{\boldmath Testing the robustness of the BAO determination in the presence of massive neutrinos}

\author[a,b]{Adriana Nadal-Matosas}
\author[a,b]{Héctor Gil-Marín,}
\author[a,c]{Licia Verde}

\affiliation[a]{Institut de Ciències del Cosmos (ICCUB), Universitat de Barcelona (UB), c.  Martí i Franquès, 1, 08028 Barcelona, Spain.}
\affiliation[b]{Institut d'Estudis Espacials de Catalunya (IEEC), Edifici RDIT, Campus UPC, 08860 Castelldefels (Barcelona), Spain}
\affiliation[c]{Institució Catalana de Recerca i Estudis Avançats (ICREA), Pg. Lluís Companys 23, E-08010 Barcelona, Spain}

\emailAdd{anadal@icc.ub.edu}
\emailAdd{hectorgil@icc.ub.edu}
\emailAdd{liciaverde@icc.ub.edu}

\abstract{We study the robustness of the Baryon Acoustic Oscillation (BAO) feature \CH{in the large-scale structure} in the presence of massive neutrinos. In the standard BAO analysis pipeline a reference cosmological model is assumed to boost the BAO peak through the so-called reconstruction technique and in the modelling of the BAO feature to extract the cosmological information. State-of-the art spectroscopic BAO measurements, such as the Dark Energy Spectroscopic Instrument claim an aggregate precision of 0.52$\%$ on the BAO scale, with a systematic error of 0.1$\%$ associated to the assumption of a reference cosmology when measuring and analyzing the BAO feature. While the systematic effect induced by this arbitrary choice of fiducial cosmology has been studied for a wide range of $\Lambda$CDM-like models, it has not yet been tested for reference cosmologies with massive neutrinos with the precision afforded by next generation surveys. In this context, we employ the \textsc{Quijote} high-resolution dark-matter simulations with haloes above a mass of $M\sim 2\times10^{13} h^{-1}M_{\odot}$, with different values for the total sum of neutrinos masses, $\sum m_\nu[{\rm eV}] = 0,\, 0.1,\, 0,2,\, 0.4$ to study and quantify the impact of the pipeline's built-in assumption of massless neutrinos on the measurement of the BAO signal, with a special focus on the BAO reconstruction technique. We determine that any additional systematic bias introduced by the assumption of massless neutrinos is no greater than $0.1\%$ ($0.2\%$) for the isotropic (anisotropic) measurement. We expect these conclusions also hold for galaxies provided  that neutrino properties do not alter the galaxy-halo connection.}

\begin{document}
\maketitle
\flushbottom

\section{Introduction}
\label{Sec:intro}

The Baryon Acoustic Oscillations (BAO) are a statistical pattern of fluctuations in the distribution of matter in the Universe and represent  one of the main probes of cosmology today, as they make it possible to infer the expansion history of the Universe through galaxy clustering in a very robust way. These oscillations have their origin in the pressure waves in the baryon-photon plasma before recombination. After recombination a characteristic distance scale, the sound horizon at radiation drag epoch $r_\text{drag}$, is imprinted in the statistical distribution of baryons and eventually also of the dark matter. The BAO feature can be identified in the distribution of galaxies observed at later epochs and can be used to map the Universe's expansion history and constrain cosmological parameters when the absolute sound horizon scale at drag epoch is known \cite{Eisenstein_2007_BAO}. In addition, even when the sound horizon scale is not known, the relative BAO measurements across and along the line-of-sight, and the relative BAO measurements at different late-time epochs, can be exploited to measure the unnormalized expansion history of the universe through galaxy clustering
\cite{alametal17,alametal21,wigglez}. 
The BAO method is now reaching high precision due to a large amount of spectroscopic redshift galaxy data and high accuracy thanks to the development of sophisticated pipelines, which include reconstruction techniques. These techniques allow the galaxy bulk velocity flows to be removed thus enhancing the BAO feature and removing its non-linear shift \cite{blasetal}. This makes the galaxy BAO one of the most robust and precise techniques to determine the cosmological expansion history, with a  $0.52\%$ aggregate statistical precision from state-of-the-art data \cite{DESI_BAO}. However, systematic errors that were considered sub-dominant to date, are now starting to potentially limit the accuracy of the BAO technique. One of them is the assumption of the reference or fiducial cosmology used to 1) construct the galaxy catalogue in term of comoving distances, 2) perform the BAO reconstruction, where the value of the ratio $f/b_1$, between the logarithmic growth rate parameter $f$ and the linear galaxy bias $b_1$ must be assumed, and 3) perform the BAO fit by assuming a fiducial power spectrum template.  

Typically, the fiducial or reference cosmology corresponds to a flat $\Lambda$CDM universe. An important concern is whether the arbitrary choice of this reference model has an impact on the cosmology inferred from BAO.
Several studies have quantified the impact of the arbitrary choice of the reference cosmology on the BAO analysis, not only within the vanilla $\Lambda$CDM model, but also for its most popular extensions \cite{assumption_impact,systematics,DESI_fiducial_cosmo_systematics,sanzwuhletal24}.
It has been found that the  systematic error floor arising from the assumption of an incorrect reference cosmology is of the order of $0.1\%$ for reasonable cosmology choices \cite{DESI_fiducial_cosmo_systematics}. This bias is safely below the statistical error budget.

But all these tests have mostly assumed that neutrinos are massless, even though recent experimental findings on neutrino flavour oscillations support the existence of massive neutrinos, setting the lower bound for the lightest neutrino specie to $m_{\nu}>0.048$ eV \cite{neutrino_oscill_mass}. Neutrino oscillations experiments can determine the neutrino's mass differences, but are not sensitive to the absolute mass scale and leave the upper bound mass unconstrained. Results from the Karlsruhe Tritium Neutrino (KATRIN)\footnote{\href{https://www.katrin.kit.edu}{https://www.katrin.kit.edu}} experiment set the upper bound for the total effective mass of neutrinos, $M_{\nu}=\sum_i m_i$, to $M_{\nu}<0.45\,{\rm eV}$  at 90$\%$ confidence level \cite{KATRIN}.

The tightest constraint for the upper bound of the total mass of neutrinos comes from cosmology. Massive neutrinos play a significant role in the history of the Universe and in the formation of structures. In particular, due to their free-streaming, they smooth out the inhomogeneities in the universe and affect the growth of cosmic clustering. As a consequence, the observable matter density power spectrum is damped on small scales  \cite{Massive_neutrinos_review}. 
Although this effect is expected to be small, their signature on the large-scale structure (LSS) can be measured and used to constrain their masses.
The latest result of the Dark Energy Spectroscopic Instrument (DESI) BAO in combination with CMB constrained the sum of neutrino masses to $M_{\nu}<0.072\,{\rm eV}$ at 95$\%$ confidence level\footnote{Under the assumption of a $\Lambda$CDM model.}  \cite{DESI_cosmology}. 
In this range of masses, the effect neutrinos have on the power spectrum ranges from a few percent in the lower limit of mass to up to 10 - 20$\%$ in the higher limit  \cite{Massive_neutrinos_review}.
Thus, in the era of high-precision cosmology, it is now unavoidable to take the added effect of massive neutrinos into account, putting the topic at the center of cosmological modelling. \CH{To do so, cosmological simulations including massive neutrinos (see for eg., Refs. \cite{Quijote,Carbone_2016,Liu_2018,Schaye_2023,10.1093/mnras/stad1657}) are employed to test the consistency of the state-of-the-art modelling of the LSS.}

Our goal is to test the robustness of the standard pipeline employed to measure the BAO feature against assumptions about neutrino masses in the reference cosmology. Specifically, we test the effect of assuming an incorrect cosmology during the reconstruction process and at the template used for BAO modelling. 
The paper is organized as follows. In Section \ref{Sec:Methodology} we present the employed simulations and the methodology followed in this study, results are presented in Section \ref{Sec:Results} and we conclude in Section \ref{Sec:Conclusions}.

\section{Methodology}
\label{Sec:Methodology}
We aim to test the dependence of the measured BAO constraints on the fiducial cosmology assumed by the standard pipeline. In order to do so, we analyse the BAO feature in Fourier space (i.e., in the power spectrum) using four \textsc{Quijote} simulations with different sum of neutrino masses $M_\nu$. We also test the impact of two different reconstruction algorithms, the Iterative Fast Fourier Transform (IFFT) algorithm \cite{Recon_Fourier} and the Multi Grid Reconstruction (MGR) algorithm  \cite{Eisenstein_2007_Reconstruction}, with two type of conventions (RecSym and RecIso, see below) for how the redshift space distortions (RSD) are treated.

We extract the BAO feature using the usual modelling in Fourier space, through a linear template of the oscillation feature in the power spectrum, damped with a non-linear parameter, and where the broadband power is described through a polynomial function, previously employed to measure the BAO in BOSS and eBOSS data (see for eg., Refs. \cite{BAO_eBOSS,BAO_sdss,BAO_boss} for further description). When analyzing the BAO, we adopt two different assumptions for building the template, one cosmology consistent the \textsc{Planck 2018} best fit \cite{Planck2018}, and the other constructed from the true underlying cosmology of the simulation (see section \ref{sec:quijote_sims}).

\subsection{Simulations}
\label{sec:quijote_sims}
We use the \textsc{Quijote} suite\footnote{\href{https://quijote-simulations.readthedocs.io/}{https://quijote-simulations.readthedocs.io/}} \cite{Quijote}, consisting of 500 independent realizations of a N-body simulations of a cube of 1($h^{-1}$Gpc)$^3$ volume with periodic boundary conditions, and containing $512^3$ cold dark matter particles (plus $512^3$ neutrino particles for simulations with massive neutrinos). 
For each of the cosmologies we consider, we use the dark matter halo catalogs of the $z=0.5$ snapshot of the 500 realizations. (see Table~\ref{tab:cosmologies} below). This choice for the redshift snapshot is motivated so that it matches as closely as possible not only the  effective redshift of the current galaxy surveys but also much of the expected effective redshift for forthcoming surveys.
Dark matter halos are identified using the Friends-of-Friends (FoF) algorithm, and only saved when containing at least 20 cold dark matter particles, which makes the \CH{halo mass cut $M\sim 2\times10^{13} h^{-1}M_{\odot}$}. \CH{The resulting mean halo number density is $\Bar{n}\sim 5.1 \times 10^{-5} (h/Mpc)^{-3}$.}
The cosmological parameters that are common among all simulations are the following: the matter and baryonic energy densities, $\Omega_m=0.3175$ and $\Omega_b=0.049$, the dark energy density is $\Omega_{\Lambda}=1-\Omega_m-\Omega_b-\Omega_{\nu}$, with $\Omega_{\nu}$ the neutrino energy density, and its equation parameter $w=-1$. This sets the value of the curvature to zero in all cases. The value for the reduced Hubble constant is $h=0.6711$. Finally, the amplitude of matter fluctuations is such that $\sigma_8=0.834$, and the spectral index is set to be $n_s=0.9624$.
Table~\ref{tab:cosmologies} displays the parameters that vary across the 4 simulations under consideration.
For comparison, the fiducial cosmology consistent with \textsc{Planck 2018} \cite{Planck2018} has the following values for parameters $\Omega_m=0.3138$ and $\Omega_b=0.0493$, $h=0.6736$, and a massive neutrino species with $\Omega_\nu= 0.0014$, corresponding to a sum of the neutrino mass of $M_{\nu}=0.05$ eV.

\begin{table}[h]
\centering
\begin{tabular}{|c|c|c|c|c|c|}
\hline
Simulation Name & $M_{\nu}$ (eV) & $r_{\rm drag}$ (Mpc) & $N^{1/3}_{\nu}$ & $f$   & $b_1$ \\ \hline\hline
zero mass        & 0.0            & 147.20               & 0               & 0.764 & 1.95  \\ \hline
low mass        & 0.1            & 147.48               & 512             & 0.762 & 1.94  \\ \hline
medium mass     & 0.2            & 147.75               & 512             & 0.761 & 1.92  \\ \hline
high mass       & 0.4            & 148.25               & 512             & 0.757 & 1.91  \\ \hline
\end{tabular}
\caption{List of the considered cosmological simulations from the \textsc{Quijote} suite; $b_1$ is estimated from the ratio of the halo-halo power spectrum to the matter-matter power spectrum at large scales.}
\label{tab:cosmologies}
\end{table}

\subsection{Power Spectrum estimator}
We obtain the first two power spectrum multipoles for each simulation using \textsc{Rustico}\footnote{\href{https://github.com/hectorgil/Rustico}{https://github.com/hectorgil/Rustico}}.
We take the mean over all 500 realizations as our data-vector to perform all the systematic tests. By doing so, we obtain the power spectrum whose statistical precision is equivalent to a volume of $500\,(h^{-1}{
\rm Gpc})^3$. 

\begin{figure}[ht]
    \centering
    \includegraphics[width=0.7\linewidth]{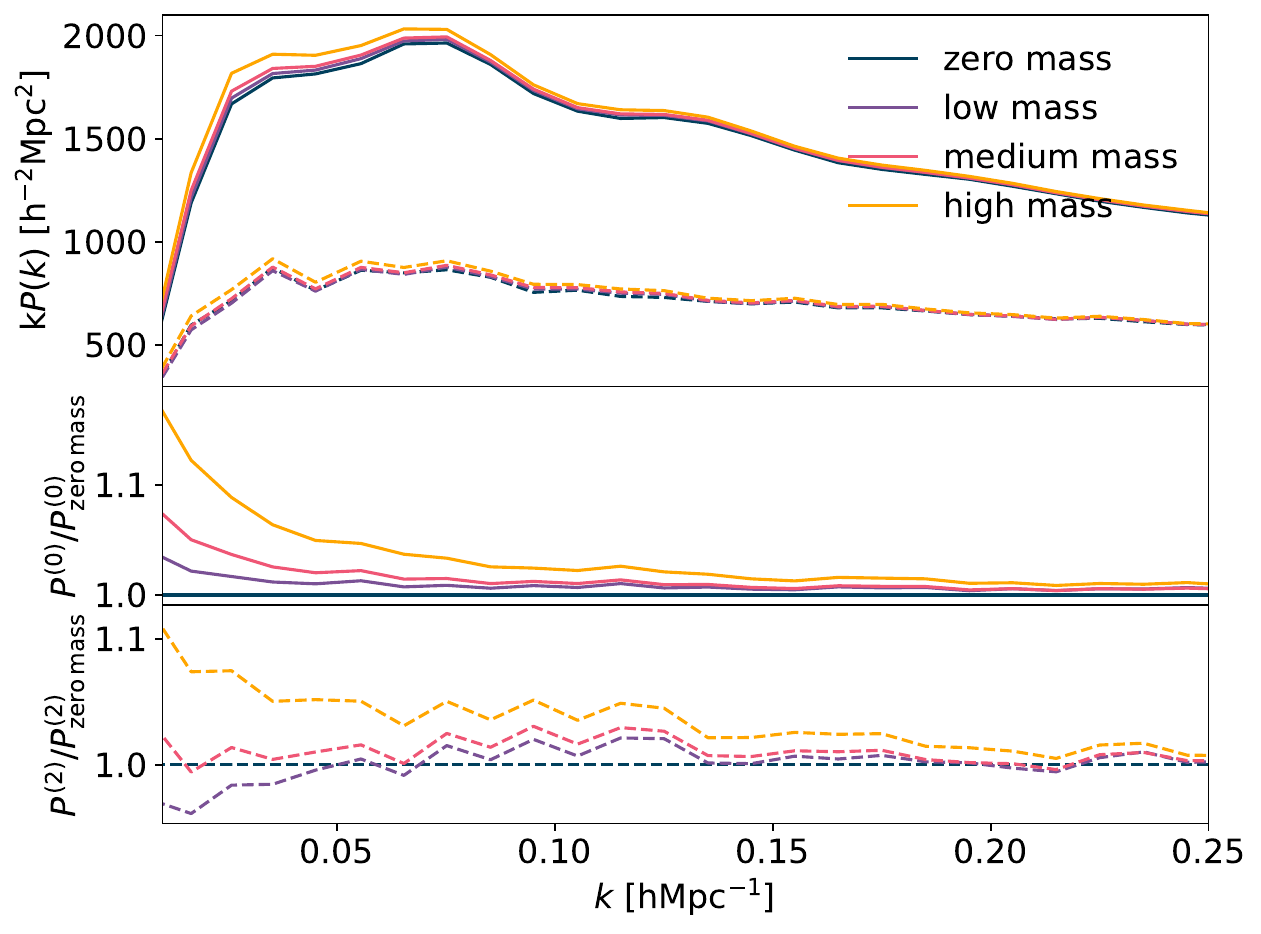}
    \caption{Power spectrum multipoles of the considered cosmological simulations from the \textsc{Quijote} suite, which are described in the text and in Table \ref{tab:cosmologies}. The first panel shows measured the monopole (solid lines) and quadrupoles (dashed lines) for a volume of 500($h^{-1}$Gpc)$^3$. The second and third panel show the ratio of the multipoles relative to the cosmological simulation without massive neutrinos.}
    \label{fig:pk_cosmo}
\end{figure}

\CH{Figure \ref{fig:pk_cosmo} shows in the fist panel the measured monopole and quadrupole corresponding to a volume of 500($h^{-1}$Gpc)$^3$ for the considered cosmological simulations described in \ref{sec:quijote_sims}. It can be observed that the oscillatory feature corresponding to the BAO remains largely unaffected by the presence of massive neutrinos. The second and third panel show, for the monopole and the quadupole respectively, the ratio of the measured multipole relative to the cosmological simulation without massive neutrinos. The clustering amplitude increases at large scales with the mass of neutrinos due to an increase of $\Omega_m$. The damping of the power spectrum at small-scales (large values of $k$) due to neutrino free-streaming is also observed.}

The reconstruction technique relies on a set of random particles that trace the geometry of the survey to evaluate the average galaxy density and estimate the overdensities in the simulations. We generate random catalogues of uniformly distributed objects for all 500 realizations. These random catalogues have 20 times more objects than the halo catalogues to minimize the shot-noise introduced in the measurement. 
\subsection{Reconstruction}
We test the potential systematic biases in the BAO signature related to the assumption of an  incorrect cosmology during the process of density reconstruction. In order to do so, we apply two different assumed cosmologies into two reconstruction algorithms with two different conventions on how the RSD are treated to the data and compare the resulting BAO signal.
We use the \textsc{pyrecon}\footnote{\href{https://github.com/cosmodesi/pyrecon}{https://github.com/cosmodesi/pyrecon}} package, which is an implementation of different algorithms and conventions for BAO reconstruction made by the DESI collaboration.
We make use of the MGR algorithm and the IFFT algorithm. \CH{The first is the lowest-order algorithm presented in \cite{Eisenstein_2007_Reconstruction}, and calculates the Zel'dovich displacements of the halos based on finite differences of the overdensity field on a grid. The second algorithm is introduced in \cite{Recon_Fourier} and follows an iterative scheme to estimate the real-space overdensities from the redshift-space overdensities, and thus the halo displacements, based on using inverse Fast Fourier Transforms. A more extensive comparison among the two can be found in \cite{recon_algorithms}.}

When it comes to the RSD treatment, we focus on two conventions. 
The RecIso convention has been used by BOSS and eBOSS collaborations \cite{alametal17,alametal21}, and consists on removing the RSD from the galaxy catalog by adding a displacement along the line of sight proportional to the velocity field causing the non-linear bulk flow. Since RSD arise from this same velocity field, this shift results in a more isotropically-clustered sample after reconstruction. 
In the RecSym convention, introduced in \cite{White2015} and used in DESI \cite{DESI_recon}, this displacement is applied both in the galaxy and random catalogues, restoring the clustering power at large scales and preserving RSD.

\begin{figure}[ht]
    \centering
    \includegraphics[width=0.8\linewidth]{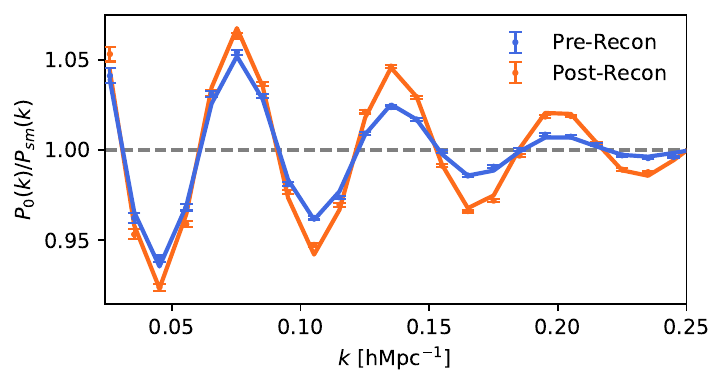}
    \caption{The isolated BAO feature in the monopole of the power spectrum before (blue) and after reconstruction (orange), obtained by normalizing the signal by $P_{\rm sm}$, the smooth broadband of power spectrum. The dots correspond to the measured power spectrum and the solid lines are the best fit BAO models.}
    \label{fig:pk_recon}
\end{figure}

We reconstruct the 500 independent realizations, for each of the four considered $M_\nu$ values, 
using both algorithms and conventions. Our baseline analysis follows \cite{BAO_eBOSS}, we set the growth rate to $f(z=0.5)=0.82$ and a galaxy bias value of $b_1=2.3$, and use a smoothing scale of 15 $h^{-1}$ Mpc. To study the impact of this choice of parameters, we repeat the reconstruction using the computed values of $f(z=0.5)$, calculated following $f(z)=\Omega_m^\gamma(z)$ with $\gamma=6/11$, and $b_1$, estimated as the ratio of the halo-halo power spectrum to the matter-matter power spectrum amplitude at large scales. For each simulation, the computed values can be found in Table \ref{tab:cosmologies}. 
\CH{Figure \ref{fig:pk_recon} shows the BAO feature pre-reconstruction and post-reconstruction. We have normalized both the signal and the best fit models by the smooth broadband to isolate the oscillations. It can be observed that the signal recovered after reconstruction is enhanced significantly, resulting in a more robust measurement of the BAO.}

\subsection{Modelling the BAO}
\label{sec:modelling_BAO}
The BAO in the radial and tangential directions provide measurements of the Hubble parameter and angular distance respectively.
The dilation parameters, $\alpha_\parallel,\alpha_\perp$, measure the shift of the BAO in the data (computed by assuming a reference model) with respect to a fiducial template model,
\begin{equation}
    \label{eq:def_alphes}
   \alpha_{\parallel} (z)=\frac{D_H(z)r_{\text{drag}}^{\text{ref}}}{D_H^{\text{ref}}(z)r_{\text{drag}}} \quad;\quad  \alpha_{\perp}(z)=\frac{D_M(z)r_{\text{drag}}^{\text{ref}}}{D_M^{\text{ref}}(z)r_{\text{drag}}}
\end{equation}
where $D_H(z)=c/H(z)$, $H(z)$ is the Hubble expansion parameter, $r_{\text{drag}}$ the comoving sound horizon at $z=z_{\text{drag}}$, and $D_M(z)$ is the comoving angular diameter distance at redshift $z$. 
The superscript "ref" refers to the values corresponding to the reference cosmologies used in the analysis. 
For $D_M^{\text{ref}}$ and $D_H^{\text{ref}}$, this cosmology is the one used to convert redshifts into distances. 
Whereas the reference $r_{\text{drag}}$ corresponds to the cosmology assumed in the underlying template used to model the BAO. 
Since our catalogs are cubes in the true underlying comoving space there is no redshift-to-distance conversion effect and we are only testing for the effect of $r_{\text{drag}}^{\text{ref}}/r_{\text{drag}}$.
In BAO analyses, any deviation from ${\alpha_\parallel}/{\alpha_\perp}=1$ in the dilation parameters ratio, is associated with an anisotropy introduced by the reference cosmology used for redshift-to-distance conversion at the catalog level. The emerging of this anisotropy is known as the Alcock-Paczynski (AP) effect \cite{AP_effect}. In our analysis, since the distances are not distorted by the fiducial cosmology choice, we expect this ratio to be unity by construction. Any deviation from unity is then a systematic error.

A common reparametrization of the BAO dilation parameters consists in an isotropic parameter, $\alpha_{\text{iso}}=\alpha_{\parallel}^{1/3}\alpha_{\perp}^{2/3}$, and 
an anisotropic one, $\alpha_{\rm AP}=\alpha_{\parallel}/\alpha_{\perp}$.
The AP effect and the isotropic dilation distort the wavenumbers along and across the line of sight giving the apparent wavenumbers $k_\parallel$ and $k_\perp$, which are related to the true ones, $k_\parallel'$ and $k_\perp'$, following $k_\parallel'=k_\parallel\alpha_\parallel$ and $k_\perp'=k_\perp\alpha_\perp$. In terms of the absolute wavenumber, $k=\sqrt{k_\parallel^2+k_\perp^2}$, and the cosine of the angle between the wavenumber vector and the line-of-sight (LOS) direction, $\mu$,

\begin{align}
    k'=\frac{k}{\alpha_{\perp}}\left\{1+\mu^2\left[\left(\frac{\alpha_{\parallel}}{\alpha_{\perp}}\right)^2-1\right]\right\}^{1/2}
    \quad;\quad 
    \mu'=\mu\frac{\alpha_{\perp}}{\alpha_{\parallel}}\left\{1+\mu^2\left[\left(\frac{\alpha_{\parallel}}{\alpha_{\perp}}\right)^2-1\right]\right\}^{-1/2}.
\end{align}
To model the power spectrum we follow the approach adopted in eBOSS \cite{BAO_eBOSS},
\begin{equation}
    P(k,\mu)=B(1+R\beta\mu^2)^2P^{(\rm sm)}_{\text{lin}}(k)\left\{1+[\mathcal{O}_{\text{lin}}(k)-1] \times e^{-k^2[\mu^2\Sigma_{\parallel}^2+(1-\mu^2)\Sigma_{\perp}^2]/2}\right\}.
    \label{eq:PS}
\end{equation}
Here, $P_{\text{lin}}$ is the linear power spectrum and $\mathcal{O}_{\text{lin}}\equiv P_{\text{lin}}/P^{(\text{sm})}_{\text{lin}}$ is the linear BAO template, where $P^{(\text{sm})}_{\text{lin}}$ is a smoothed power spectrum with no BAO signal.
$B$ is identified with the linear bias squared, $b_1^2$, $\beta$ is the redshift space distortion parameter, and $R$ accounts for the redshift-space distortion suppression due to reconstruction. 
For reconstruction in the RecSym convention, since RSD are conserved, we set $R=1$. For the RecIso convention the reconstructed field presents RSD suppression in large scales, so we set $R=1-\exp(-(k\Sigma_s)^2/2)$, where $\Sigma_s$ is the smoothing scale, which takes the value of $15\,{\rm Mpc}h^{-1}$ for all the cases presented in this paper. The damping of the BAO due to bulk motions of galaxies is described by the parameters $\Sigma_{\parallel}$ and $\Sigma_{\perp}$, along and across the LOS respectively.

We fit $\alpha_{\parallel}$ and $\alpha_{\perp}$, as well as $B$, $\beta$, $\Sigma_\parallel$, $\Sigma_\perp$ and 5 coefficients per multipole for the polynomial describing the broadband of the power spectrum. We do so by exploring the likelihood surface using \textsc{Brass}\footnote{\href{https://github.com/hectorgil/Brass}{https://github.com/hectorgil/Brass}}, an MCMC solver based on the Metropolis–Hasting algorithm with a proposal covariance and a Gelman-Rubin convergence criteria of $R_{\rm GR}-1\leq0.01$. The measurements on $\alpha_{\parallel}$ and $\alpha_{\perp}$ are always marginalised over the rest of nuisance parameters, including $B$, $\beta$, $\Sigma_{\parallel,\perp}$ and the 5 coefficients per multipole.
We follow the scheme in two ways, to test the bias introduced by the assumption of an incorrect cosmology in the underlying template. First, we make use of the publicly available $P^{(\text{sm})}_{\text{lin}}$ and $\mathcal{O}_{\text{lin}}$ used in \cite{BAO_eBOSS}. They are generated following the \textsc{Planck 2018} \cite{Planck2018} reference cosmology with $r_{\text{drag}}^{\text{ref}}=147.78$ Mpc and $h=0.677$. Alternatively, we generate a set of $P^{(\text{sm})}_{\text{lin}}$ and $\mathcal{O}_{\text{lin}}$ corresponding to each of the four \textsc{Quijote} cosmological simulations obtaining $P_{\text{lin}}$ using \textsc{class} \cite{Class11} and separating the broadband to from the oscillatory feature following the methodology described in \cite{Kirkbyetal13}.

In the absence of any systematic effect the recovered values of $\alpha_{\parallel}$ and $\alpha_{\perp}$ are expected to be consistent with the values given by equation \ref{eq:def_alphes}. Omitting the redshift-to-distance conversion effect, which our catalogs do not display, the expected values are then, 
\begin{equation}
\alpha_{\parallel,\perp}^{\text{expected}}=r_{\text{drag}}^{\text{ref}}/r_{\text{drag}}.
\end{equation}
For the alternative parametrization this reads as,
\begin{equation}
\alpha_{\text{iso}}^{\text{expected}}=r_{\text{drag}}^{\text{ref}}/r_{\text{drag}}; \quad\quad \alpha_{\rm AP}^{\text{expected}}=1. 
\end{equation}
Therefore, we quantify the bias as the difference
\begin{equation}
    \Delta\alpha= \alpha^{\text{measured}}- \alpha^{\text{expected}}.
\end{equation}

\section{Results}
\label{Sec:Results}
The evaluation of the robustness on the choice of reconstruction algorithm, convention, parameters and underlying template of the fit is presented here. Additionally, to evaluate if these systematics represent a limitation for the current or future DESI BAO measurements, we compare the resulting systematic error budget  to that of DESI presented in \cite{DESI_fiducial_cosmo_systematics}. 

We choose our baseline analysis to follow the IFFT reconstruction algorithm, in the RecSym convention, with $f$ and $b_1$ consistent with \textsc{Planck 2018} as employed in \cite{BAO_eBOSS} and fitting the BAO signal employing the fiducial template of the \textsc{Planck 2018} cosmology. 
We summarize our baseline analysis measurement in Table~\ref{tab:baseline} where the reported errors correspond to the usual $68.27\%$ confidence interval ($1\sigma$). Hereafter, we adopt the convention that a systematic shift can only be said to be detected if it is above $95.45\%$ confidence level ($2\sigma$).

The recovered shifts are compatible with $\Delta\alpha=0$ within $2\sigma$, hence we can only place  upper limits to the shifts  of  $0.23\%$ ($0.14\%$) for $\Delta\alpha_\parallel$ ($\Delta\alpha_\perp$). These are well below the error budget in current surveys \cite{DESI_systematics}.

\begin{table}[h]
\centering
\begin{tabular}{|c|c|c|c|c|}
\hline
Simulation Name & $\Delta\alpha_\parallel$ [$\%$] & $\Delta\alpha_\perp$ [$\%$] & $\Delta\alpha_{\rm iso}$ [$\%$] & $\Delta\alpha_{\rm AP}$ [$\%$] \\ \hline\hline
zero mass       & -0.23 $\pm$ 0.10                    & -0.06 $\pm$ 0.07                & -0.12 $\pm$ 0.05                 & -0.16 $\pm$ 0.15                \\ \hline
low mass        & -0.10 $\pm$ 0.11                    & 0.01 $\pm$ 0.07                 & -0.02 $\pm$ 0.05                 & -0.11 $\pm$ 0.15                \\ \hline
medium mass     & -0.13 $\pm$ 0.13                    & 0.11 $\pm$ 0.07                 & 0.02 $\pm$ 0.05                  & -0.24 $\pm$ 0.18                \\ \hline
high mass       & -0.05 $\pm$ 0.14                    & 0.14 $\pm$ 0.07                 & 0.07 $\pm$ 0.05                  & -0.18 $\pm$ 0.18                \\ \hline 
\end{tabular}
\caption{Baseline analysis measurements of $\Delta\alpha_\parallel$, $\Delta\alpha_\perp$, $\Delta\alpha_{\rm iso}$ and $\Delta\alpha_{\rm AP}$ for all simulations with varying neutrino mass. We report the best fit value and the errors correspond to $1\sigma$ statistical errors for the considered volume of $500\,( h^{-1}{\rm Gpc})^3$.}
\label{tab:baseline}
\end{table}

\begin{figure}[ht]
    \centering
    \includegraphics[width=0.85\linewidth]{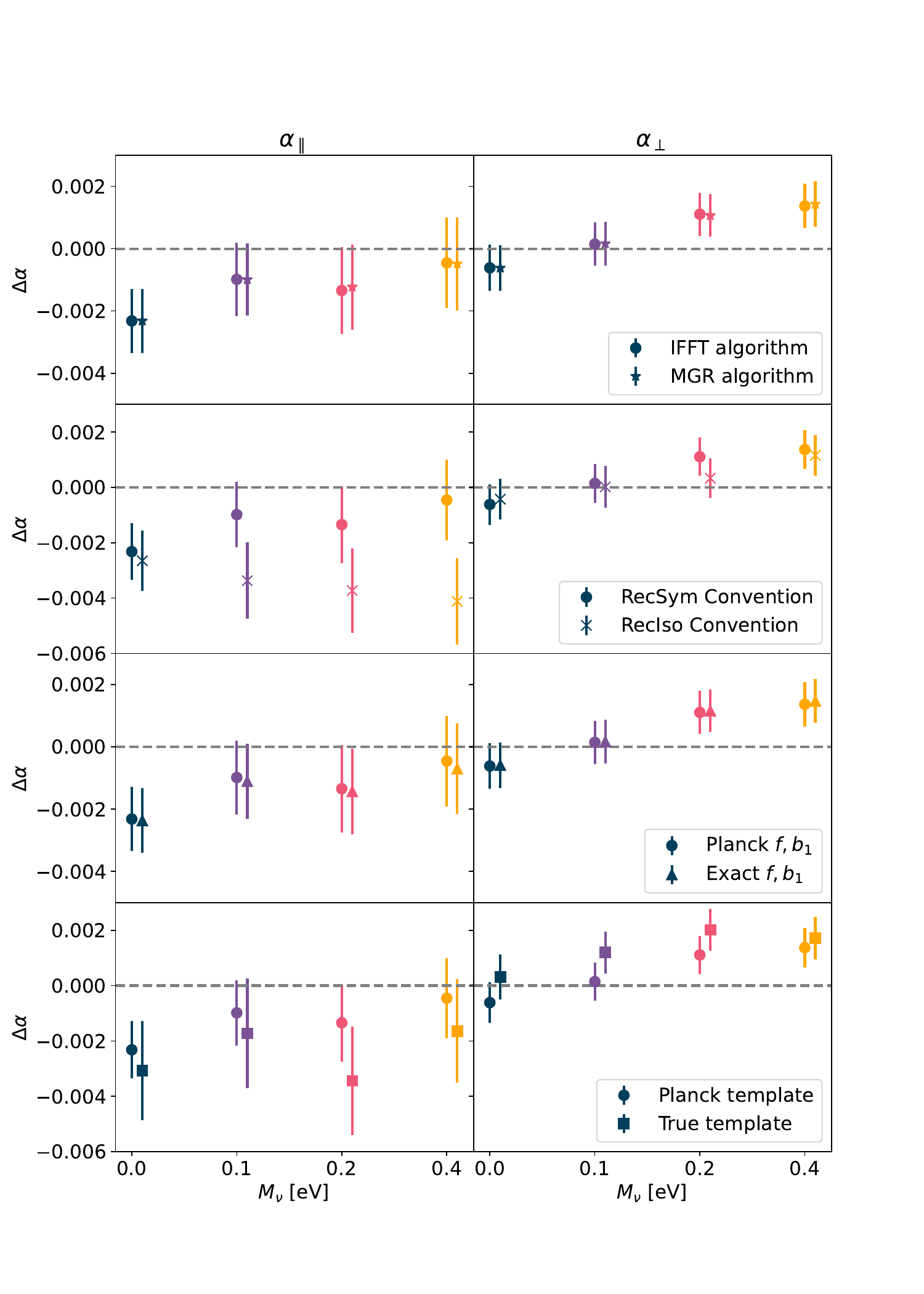}
    \caption{Comparison of the best-fit measurements of $\Delta\alpha_{\parallel}$ (left panels) and $\Delta\alpha_{\perp}$ (right panels) obtained from isolating each assumption to our baseline analysis. We compare the choice of algorithms (top panels), conventions for RSD treatment (second panels), choice of parameters for reconstruction (third panels) and underlying template used during the fit (bottom panels). Round markers always correspond to our baseline analysis, corresponding to reconstruction with the IFFT algorithm, the RecSym convention and \textsc{Planck 2018}'s $f$ and $b_1$ and fitting the signal with the underlying template generated from \textsc{Planck 2018}.}
    \label{fig:paraperp}
\end{figure}

We test each choice/assumption on the reconstruction algorithm, RSD treatement, and fiducial model and template by isolating it and report the effect on the measured values of $\Delta\alpha_{\parallel,\perp}$ in Figure~\ref{fig:paraperp}. 
Left panels show $\Delta\alpha_{\parallel}$, while right panels show $\Delta\alpha_{\perp}$. Each colour corresponds to different values for the mass of neutrinos in the simulations used. 
We show the best fit of our baseline analysis and each of our set of tests. The errors are obtained from the likelihood study corresponding to $1\sigma$ of the combined volume of 500($h^{-1}$Gpc)$^3$ and only include the statistical error.
The panels at the top display the comparison between the adopted reconstruction algorithms. Both algorithms give consistent results, with a relative shift lower than $0.01\%$ for all neutrino masses considered. This result is in agreement with the findings in \cite{recon_algorithms}. 
When it comes to RSD treatment, the second panels display the comparison between the RecSym and RecIso conventions. Here we observe a difference in the measured $\alpha_\parallel$ that increases with the mass of neutrinos. This difference becomes statistically relevant at a volume of 500($h^{-1}$Gpc)$^3$ for $M_\nu=0.4$eV, where the relative shift is $0.36\%$ (3$\sigma$) for $\alpha_\parallel$. 
The performance of $\alpha_\perp$ is consistent for both conventions through the neutrino mass range considered.
The third panels show the comparison between reconstructing with reference or exact values of growth rate and linear bias. We find the measurement is not sensitive to this choice, with their relative shift being always lower than $0.01\%$ for all neutrino masses considered. 
The bottom panels show the measured alphas obtained with the \textsc{Planck 2018} and true underlying cosmology templates in the fitting. It is interesting to note that using the true templates yield larger statistical error-bars for all neutrino masses considered. This is discussed further in the Appendix \ref{sec:appendix_A}. The measurements for both templates are consistent, with a maximum relative shift of $0.2\%$ for $\alpha_\parallel$ in the medium mass simulation, well within the statistical error. The performance for $\alpha_\perp$ is also consistent for both templates through the neutrino mass range considered.
The systematic shifts induced by the choice of the fiducial neutrino mass are at the $0.2\%$ level and are well below the shifts induced by different choices of the algorithm and RSD treatment convention. 

\subsection{Comparing with DESI DR1}
\begin{figure}[ht]
    \centering
    \includegraphics[width=\linewidth]{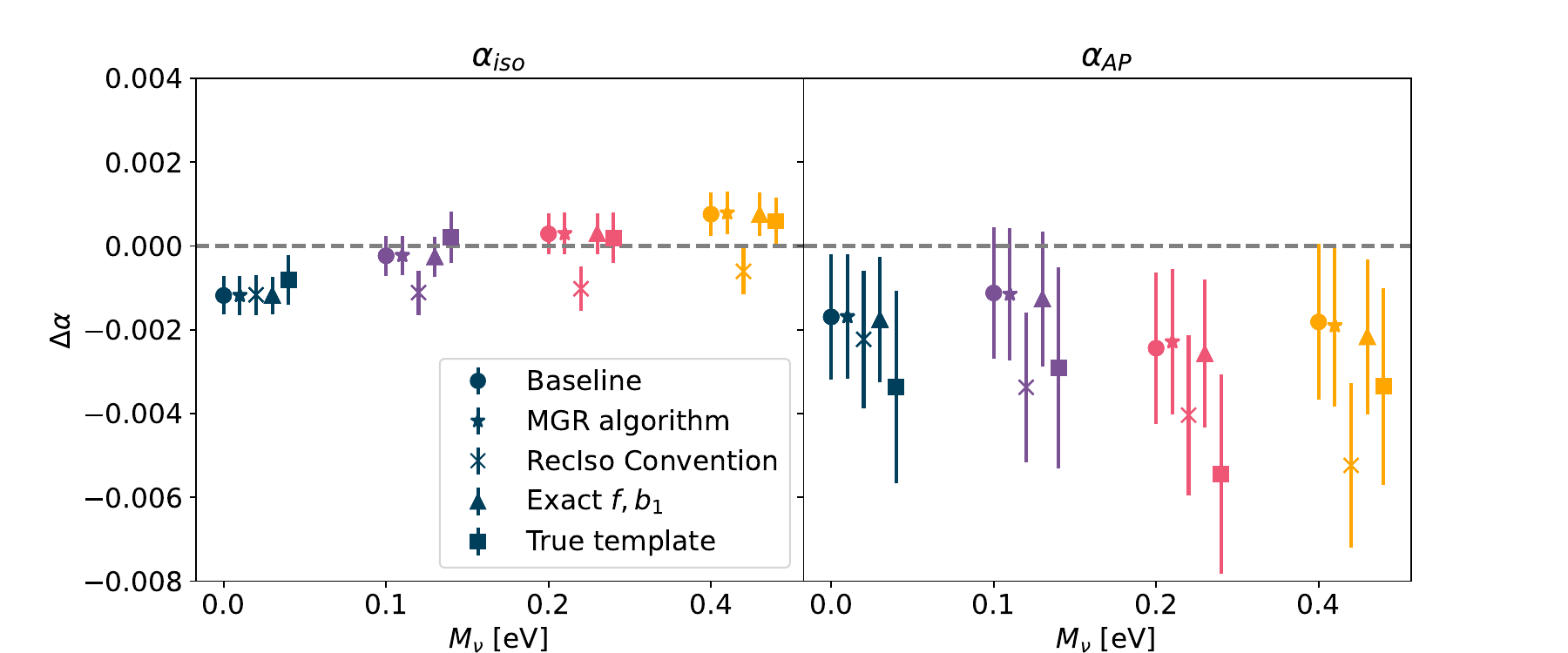}
    \caption{Comparison of the best-fit measurements of $\Delta\alpha_{\text{iso}}$ (left panel) and $\Delta\alpha_{\rm AP}$ (right panel) obtained from isolating each assumption of our baseline analysis. We compare the choice of algorithms, conventions for RSD treatment, choice of parameters for reconstruction and underlying template used during the fit. Round markers always correspond to our baseline analysis, reconstructing with the IFFT algorithm, the RecSym convention and \textsc{Planck 2018}'s $f$ and $b_1$ and fitting the signal with the underlying template generated from \textsc{Planck 2018}.}
    \label{fig:isoap}
\end{figure}

\CH{Recently the DESI Collaboration has published BAO DR1 results from galaxies and quasars \cite{DESI_BAO}, supported by a study on the theoretical and modelling systematics associated to the measurement \cite{DESI_systematics}. 
They find a combined systematic error contributions from both theory and modelling of $0.1\%$ ($0.2\%$) for $\alpha_{\text{iso}}$ ($\alpha_{\rm AP}$).
This result explores the impact of modelling choices on the BAO constraint and includes their studies of systematic errors introduced during reconstruction \cite{DESI_recon} and the dependence on the choice of fiducial cosmology \cite{DESI_fiducial_cosmo_systematics} of the measurement. 
In this section we compare their findings with our study. The halo catalogues employed in this work have similar bias properties as the LRG DESI sample ($b_1\simeq 2$). However, we do not expect that the systematic study of this work to strongly depend on this choice, and in principle could be extrapolated to other samples, such as ELG and QSO.} The results in the $\alpha_{\text{iso},\rm AP}$ parametrisation, are shown in Figure~\ref{fig:isoap}. 
For $\alpha_{\text{iso}}$ we find all our measurements are well within a $0.1\%$ shift of their expected value. We only recover a statistically significant shift of $0.13\%$ (3$\sigma$) when using the RecIso convention for $M_\nu>0.2$eV. 
For $\alpha_{\rm AP}$ we find all measurements are compatible with our baseline analysis within the statistical error of the volume considered. 
We conclude that this measurement is statistically consistent with the $0.2\%$ recovered for the baseline analysis and the total systematic error budget found by DESI for $\alpha_{\rm AP}$.

\section{Conclusions}
\label{Sec:Conclusions}

In this study, we have used the halo catalogs from the \textsc{Quijote} simulations suite to test the standard assumption of massless neutrinos through the BAO pipeline. 
We have tested the effect of assuming an incorrect mass of neutrinos during density reconstruction and during the BAO analysis. We have presented our results for $\alpha_{\parallel}$ and $\alpha_{\perp}$ from measuring relative differences between the BAO wiggle positions in the power spectrum with respect to the template.

We show that the two different reconstruction  algorithms perform in an almost identical way for all neutrino masses considered, presenting relative shifts of $<0.1\%$.
As for the two RSD treatment conventions  we find that the RecIso convention consistently performs worse than the RecSym convention thought the neutrino mass range considered, with relative shifts on $\alpha_{\parallel}$ of $0.36\%$ for the largest neutrino mass considered. The RecIso convention, although widely used in the past, has been shown to introduce non-negligible shifts in the BAO measurement due to uncontrolled long modes \cite{BAO_IR_modes}, while removing BAO information from the anisotropic clustering and being more cosmology-choice dependent \cite{DESI_systematics}. Our result is in agreement with this finding, and supports the argument that the RecSym convention is more robust against reference cosmology assumptions.
Independently of the choice of parameters and neutrino mass, we show that the bias arising from assuming incorrect values of $f$ and $b_1$ in the density reconstruction technique is always $<0.1\%$.
Finally, we have tested the measurement against the assumption of an incorrect cosmology in the underlying template during the fitting of the BAO signal. When comparing the measured values of $\alpha_{\parallel,\perp}$, the relative shift amongst templates has been found to be well below the statistical errors for a volume of 500($h^{-1}$Gpc)$^3$, independently on the mass of neutrinos. This shifts have been found to be $<0.2\%$ ($<0.1\%$) for $\alpha_{\parallel}$ ($\alpha_{\perp}$), which translate to $0.01\%$ ($0.2\%$) for $\alpha_{\text{iso}}$ ($\alpha_{\rm AP}$).

We conclude that there is no additional systematic bias related to the assumption of reference cosmologies that consider neutrinos massless throughout the standard anisotropic pipeline for measuring the BAO to $0.1\%$ ($0.2\%$) for the isotropic (anisotropic) measurement.

The contribution of this bias is comparable to the systematic error floor of the current DESI analysis \cite{DESI_systematics}. 

The template choice, however, has an effect on the size of the (statistical) error bars for the large volumes considered here, this is further discussed in Appendix~\ref{sec:appendix_A}. However, this is not a concern for volumes comparable to those of on-going surveys.
This work complements previous studies on the dependence of the BAO measurement on assumptions made throughout the density reconstruction of catalogues as well as throughout the fitting of the signal by expanding the tests to cosmologies considering massive neutrinos. It serves to evaluate in a quantitative way the robustness of the BAO analysis against modelling assumptions of massless neutrinos, and puts on robust grounds any future constrain or detection of neutrino masses from clustering of galaxy redshift surveys.
\acknowledgments
We acknowledge the support of Center of Excellence Maria de Maeztu 2020-2023's award to the ICCUB (CEX2019-000918- M funded by MCIN/AEI/10.13039/501100011033). 
ANM and HGM acknowledges support through the Leonardo program (LEO23-1-897) of the BBVA foundation. HGM acknowledges support through the program Ramón y Cajal (RYC-2021-034104) of the Spanish Ministry of Science and Innovation.  LV and HGM acknowledge the support of project PID2022-141125NB-I00 MCIN/AEI.

\appendix
\section{Dependence of the errors on the choice of template}
\label{sec:appendix_A}

\begin{figure}[ht]
    \centering
     \includegraphics[width=0.4\linewidth]{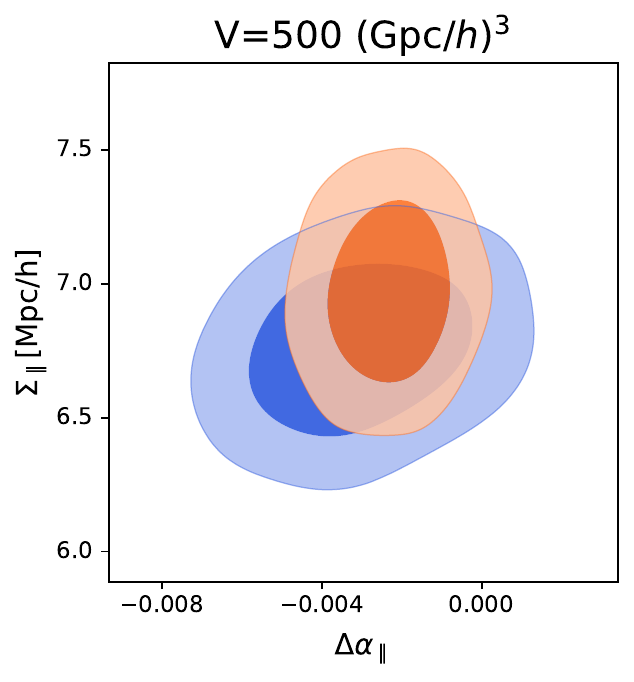}
     \includegraphics[width=0.4\linewidth]{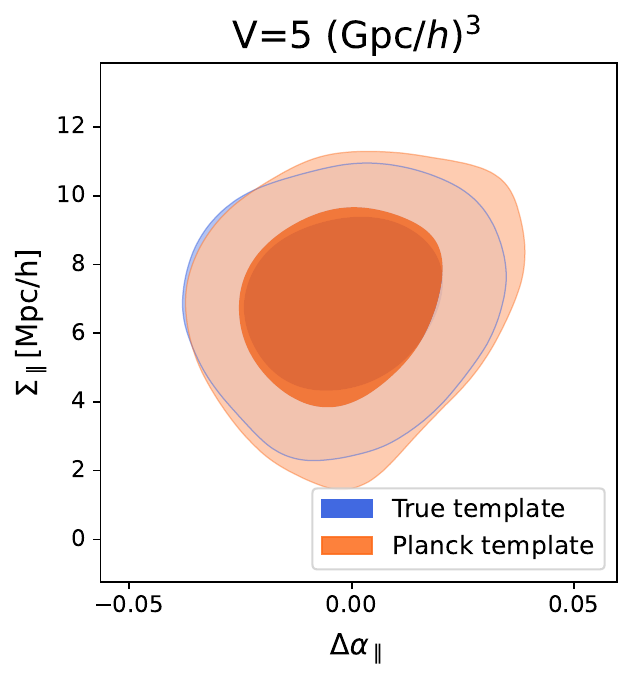}
    \caption{Left panel: Comparison of the $\Sigma_\parallel$-dependence of the $\alpha_\parallel$ measurement. The blue contour shows the $1\sigma$ and $2\sigma$ posteriors of the fit when the underlying template is constructed from the true cosmology of the simulation. The orange contour shows the posteriors of the fit when the underlying template is constructed from the \textsc{Planck 2018} cosmology. Both blue and orange contours correspond to a covariance with a volume of 500($h^{-1}$Gpc)$^3$, as in Figure~\ref{fig:paraperp}. Right panel: same as the left panel for a covariance matrix rescaled to a volume of 5($h^{-1}$Gpc)$^3$ and with a wide Gaussian prior on $\Sigma_\parallel$ of $\mathcal{N}(\mu=7,\sigma=3)$ to avoid unphysical values. The differences shown in the left panel become negligible when the size of the covariance correspond to a value close to the one of an actual galaxy survey, as displayed in the right panel.}
    \label{fig:sigma_parallel}
\end{figure}

We make a detailed investigation on the differences in the statistical precision of $\Delta\alpha_\parallel$ when using the two underlying templates reported in Figure~\ref{fig:paraperp} and \ref{fig:isoap}. The difference arises from the fact that the two templates produce --unsurprisingly-- different best fit values for some of the nuisance parameters, in particular some of the coefficients of the polynomial governing the broadband of the power spectrum and the BAO damping. The effect on the statistical error bars is particularly evident for the $\Sigma_{\parallel}$ parameter. In the left panel of Figure~\ref{fig:sigma_parallel} we compare the posteriors obtained for $\Delta\alpha_\parallel$ and $\Sigma_\parallel$ for the simulations without massive neutrinos when fitting the signal with the \textsc{Planck 2018} and true templates. We find that the increase in the statistical error for $\Delta\alpha_\parallel$ arises from the change in the degeneracy direction with $\Sigma_\parallel$. For each template the response of the fit to a variation in nuisance parameters, like $\Sigma_\parallel$, is different. This is then reflected accordingly in the associated errors. In terms of the $\alpha_{\rm iso}$ and $\alpha_{\rm AP}$, we have observed that the change in $\Sigma_\parallel$ mainly impacts the statistical errors of $\alpha_{\rm AP}$, and leaves $\alpha_{\rm iso}$ unaffected, as one could have expected from the change in errors in Figure~\ref{fig:isoap}.

In the right panel of Figure~\ref{fig:sigma_parallel} we display the same contour plot as in the left panel but for a volume of 5($h^{-1}$Gpc)$^3$, comparable to on-going galaxy redshift surveys. We observe that for such a volume this effect is undetectable, hence not a major concern.

\bibliographystyle{ieeetr}
\bibliography{references}

\begin{thebibliography}{10}

\bibitem{Eisenstein_2007_BAO}
D.~J. {Eisenstein}, H.~{Seo}, and M.~{White}, ``On the robustness of the acoustic scale in the low-redshift clustering of matter,'' {\em The Astrophysical Journal}, vol.~664, no.~2, p.~660, 2007.

\bibitem{alametal17}
S.~{Alam} {\em et~al.}, ``{The clustering of galaxies in the completed SDSS-III Baryon Oscillation Spectroscopic Survey: cosmological analysis of the DR12 galaxy sample},'' {\em \mnras}, vol.~470, no.~3, pp.~2617--2652, 2017.

\bibitem{alametal21}
S.~{Alam} {\em et~al.}, ``{Completed SDSS-IV extended Baryon Oscillation Spectroscopic Survey: Cosmological implications from two decades of spectroscopic surveys at the Apache Point Observatory},'' {\em \prd}, vol.~103, no.~8, p.~083533, 2021.

\bibitem{wigglez}
C.~{Blake} {\em et~al.}, ``{The WiggleZ Dark Energy Survey: mapping the distance-redshift relation with baryon acoustic oscillations},'' {\em \mnras}, vol.~418, no.~3, pp.~1707--1724, 2011.

\bibitem{blasetal}
D.~{Blas} {\em et~al.}, ``{Time-sliced perturbation theory II: baryon acoustic oscillations and infrared resummation},'' {\em \jcap}, vol.~2016, no.~7, p.~028, 2016.

\bibitem{DESI_BAO}
{DESI Collaboration} {\em et~al.}, ``{DESI 2024 III: Baryon Acoustic Oscillations from Galaxies and Quasars},'' {\em arXiv e-prints}, p.~arXiv:2404.03000, 2024.

\bibitem{assumption_impact}
P.~{Carter} {\em et~al.}, ``{The impact of the fiducial cosmology assumption on BAO distance scale measurements},'' {\em \mnras}, vol.~494, no.~2, pp.~2076--2089, 2020.

\bibitem{systematics}
M.~{Vargas-Maga{\~n}a} {\em et~al.}, ``{The clustering of galaxies in the completed SDSS-III Baryon Oscillation Spectroscopic Survey: theoretical systematics and Baryon Acoustic Oscillations in the galaxy correlation function},'' {\em \mnras}, vol.~477, no.~1, pp.~1153--1188, 2018.

\bibitem{DESI_fiducial_cosmo_systematics}
A.~{P{\'e}rez-Fern{\'a}ndez} {\em et~al.}, ``{Fiducial-Cosmology-dependent systematics for the DESI 2024 BAO Analysis},'' {\em arXiv e-prints}, p.~arXiv:2406.06085, 2024.

\bibitem{sanzwuhletal24}
S.~{Sanz-Wuhl} {\em et~al.}, ``{BAO cosmology in non-spatially flat background geometry from BOSS+eBOSS and lessons for future surveys},'' {\em \jcap}, vol.~2024, no.~5, p.~116, 2024.

\bibitem{neutrino_oscill_mass}
R.~L. {Workman} {\em et~al.}, ``{Review of Particle Physics},'' {\em PTEP}, vol.~2022, p.~083C01, 2022.

\bibitem{KATRIN}
T.~K. Collaboration, ``Direct neutrino-mass measurement with sub-electronvolt sensitivity,'' {\em Nat. Phys.}, vol.~18, p.~160–166, 2022.

\bibitem{Massive_neutrinos_review}
J.~{Lesgourgues} and S.~{Pastor}, ``Massive neutrinos and cosmology,'' {\em Physics Reports}, vol.~429, no.~6, pp.~307--379, 2006.

\bibitem{DESI_cosmology}
D.~Collaboration, ``{DESI 2024 VI: Cosmological Constraints from the Measurements of Baryon Acoustic Oscillations},'' {\em arXiv e-prints}, p.~arXiv:2404.03002, 2024.

\bibitem{Quijote}
F.~{Villaescusa-Navarro} {\em et~al.}, ``{The Quijote Simulations},'' {\em \apjs}, vol.~250, no.~1, p.~2, 2020.

\bibitem{Carbone_2016}
C.~{Carbone}, M.~{Petkova}, and K.~{Dolag}, ``Demnuni: Isw, rees-sciama, and weak-lensing in the presence of massive neutrinos,'' {\em \jcap}, vol.~2016, no.~07, p.~034–034, 2016.

\bibitem{Liu_2018}
J.~{Liu} and othern, ``Massivenus: cosmological massive neutrino simulations,'' {\em \jcap}, vol.~2018, no.~03, p.~049–049, 2018.

\bibitem{Schaye_2023}
J.~o. {Schaye}, ``The flamingo project: cosmological hydrodynamical simulations for large-scale structure and galaxy cluster surveys,'' {\em \mnras}, vol.~526, no.~4, p.~4978–5020, 2023.

\bibitem{10.1093/mnras/stad1657}
C.~{Hernández-Aguayo} {\em et~al.}, ``The millenniumtng project: high-precision predictions for matter clustering and halo statistics,'' {\em \mnras}, vol.~524, no.~2, pp.~2556--2578, 2023.

\bibitem{Recon_Fourier}
A.~{Burden}, W.~J. {Percival}, and C.~{Howlett}, ``{Reconstruction in Fourier space},'' {\em \mnras}, vol.~453, no.~1, pp.~456--468, 2015.

\bibitem{Eisenstein_2007_Reconstruction}
D.~J. {Eisenstein} {\em et~al.}, ``Improving cosmological distance measurements by reconstruction of the baryon acoustic peak,'' {\em The Astrophysical Journal}, vol.~664, no.~2, p.~675, 2007.

\bibitem{BAO_eBOSS}
H.~Gil-Marín {\em et~al.}, ``{The Completed SDSS-IV extended Baryon Oscillation Spectroscopic Survey: measurement of the BAO and growth rate of structure of the luminous red galaxy sample from the anisotropic power spectrum between redshifts 0.6 and 1.0},'' {\em \mnras}, vol.~498, no.~2, pp.~2492--2531, 2020.

\bibitem{BAO_sdss}
F.~{Beutler} {\em et~al.}, ``{The clustering of galaxies in the completed SDSS-III Baryon Oscillation Spectroscopic Survey: baryon acoustic oscillations in the Fourier space},'' {\em \mnras}, vol.~464, no.~3, pp.~3409--3430, 2017.

\bibitem{BAO_boss}
H.~{Seo} {\em et~al.}, ``{Modeling the reconstructed BAO in Fourier space},'' {\em \mnras}, vol.~460, no.~3, pp.~2453--2471, 2016.

\bibitem{Planck2018}
{Planck Collaboration}, ``Planck 2018 results - vi. cosmological parameters,'' {\em A\&A}, vol.~641, p.~A6, 2020.

\bibitem{recon_algorithms}
X.~{Chen} {\em et~al.}, ``{Extensive analysis of reconstruction algorithms for DESI 2024 baryon acoustic oscillations},'' {\em arXiv e-prints}, p.~arXiv:2411.19738, 2024.

\bibitem{White2015}
M.~{White}, ``{Reconstruction within the Zeldovich approximation},'' {\em \mnras}, vol.~450, no.~4, pp.~3822--3828, 2015.

\bibitem{DESI_recon}
E.~{Paillas} {\em et~al.}, ``{Optimal Reconstruction of Baryon Acoustic Oscillations for DESI 2024},'' {\em arXiv e-prints}, p.~arXiv:2404.03005, 2024.

\bibitem{AP_effect}
C.~{Alcock} and B.~{Paczynski}, ``{An evolution free test for non-zero cosmological constant},'' {\em \nat}, vol.~281, p.~358, 1979.

\bibitem{Class11}
D.~{Blas}, J.~{Lesgourgues}, and T.~{Tram}, ``The cosmic linear anisotropy solving system (class). part ii: Approximation schemes,'' {\em Journal of Cosmology and Astroparticle Physics}, vol.~2011, no.~07, p.~034, 2011.

\bibitem{Kirkbyetal13}
D.~{Kirkby} {\em et~al.}, ``Fitting methods for baryon acoustic oscillations in the lyman-$\alpha$ forest fluctuations in boss data release 9,'' {\em Journal of Cosmology and Astroparticle Physics}, vol.~2013, no.~03, p.~024, 2013.

\bibitem{DESI_systematics}
S.~{Chen} {\em et~al.}, ``{Baryon Acoustic Oscillation Theory and Modelling Systematics for the DESI 2024 results},'' {\em arXiv e-prints}, p.~arXiv:2402.14070, 2024.

\bibitem{BAO_IR_modes}
N.~{Sugiyama}, ``{Developing a Theoretical Model for the Resummation of Infrared Effects in the Post-Reconstruction Power Spectrum},'' {\em arXiv e-prints}, p.~arXiv:2402.06142, 2024.

\end{thebibliography}
\end{document}